\documentclass{mn2e}
\usepackage{amsfonts}
\usepackage{amsmath}
\usepackage{amssymb}
\usepackage{graphicx}

\newcommand{\sig}{\ensuremath{\sigma}}

\newcommand{\avg}[1]{\ensuremath{\langle \,#1\, \rangle}}
\newcommand{\etal}{et al.}

\newcommand{\DM}{\ensuremath{\Delta M}}

\newcommand{\pgal}{\ensuremath{p_{\rm gal}(L)}}
\newcommand{\Pgal}{\ensuremath{P_{\rm gal}}}
\newcommand{\Ngal}{\ensuremath{N}}
\newcommand{\psat}{\ensuremath{p_{\rm sat}(L)}}
\newcommand{\Psat}{\ensuremath{P_{\rm sat}}}
\newcommand{\Nsat}{\ensuremath{N_{\rm sat}}}

\newcommand{\Lsun}{\ensuremath{L_{\odot}}}

\newcommand{\eqn}[1]{equation~\eqref{#1}}
\newcommand{\eqns}[1]{equations~\eqref{#1}}
\newcommand{\fig}[1]{Figure~\ref{#1}}

\newcommand{\ph}[1]{\phantom{#1}}

\newcommand{\be}{\begin{equation}}
\newcommand{\ee}{\end{equation}}
\newcommand{\Cal}[1]{\ensuremath{\mathcal{#1}}}

\title[BCG and BSG luminosities]
      {The luminosities of the brightest cluster galaxies and 
       brightest satellites in SDSS groups}

\author[A. Paranjape \& R. K. Sheth]
{Aseem Paranjape$^{1}$\thanks{E-mail: aparanja@ictp.it} \& 
 Ravi K. Sheth$^{1,2}$\\ 
 $^1$ The Abdus Salam International Center for Theoretical Physics, 
      Strada Costiera 11, 34151 Trieste, Italy\\
 $^2$ Center for Particle Cosmology, University of Pennsylvania, 
      209 S. 33rd St., Philadelphia, PA 19104, USA}
\begin{document}
\pagerange{\pageref{firstpage}--\pageref{lastpage}}

\maketitle 

\label{firstpage}

\begin{abstract}
We show that the distribution of luminosities of Brightest Cluster 
Galaxies in an SDSS-based group catalog suggests that BCG luminosities 
are just the statistical extremes of the group galaxy luminosity 
function.  This latter happens to be very well approximated by the 
all-galaxy luminosity function (restricted to $M_r<-19.9$), 
provided one uses a parametrization of this function that is accurate 
at the bright end.  
A similar analysis of the luminosity distribution of the Brightest 
Satellite Galaxies suggests that they are best thought of as being 
the second brightest pick from the same luminosity distribution of 
which BCGs are the brightest.  I.e., BSGs are not the brightest of 
some universal satellite luminosity function, in contrast to what Halo 
Model analyses of the luminosity dependence of clustering suggest.  
However, we then use mark correlations to provide a novel test of these 
order statistics, showing that the hypothesis of a universal 
luminosity function (i.e. no halo mass dependence) from which the BCGs 
and BSGs are drawn is incompatible with the data, despite the fact that 
there was no hint of this in the BCG and BSG luminosity distributions 
themselves.  
We also discuss why, since extreme value statistics are explicitly a 
function of the number of draws, the consistency of BCG luminosities 
with extreme value statistics is most clearly seen if one is careful 
to perform the test at fixed group richness \Ngal.  Tests at, e.g., 
fixed total group luminosity $L_{\rm tot}$, will generally be biased and 
may lead to erroneous conclusions.  
\end{abstract}


\section{Introduction}

The study of whether or not the brightest cluster galaxy (the BCG) is
special, rather than simply being the brightest by chance, has a long 
history (Scott 1957; Schechter \& Peebles 1976; Tremaine \& Richstone 1977; 
Bhavsar \& Barrow 1985; Loh \& Strauss 2006; Lin, Ostriker \& Miller
2010, Dobos \& Csabai 2011).  
Since the BCG is usually at or close to the cluster center, it is 
plausible that its formation history was dominated by different physical 
processes compared to the other galaxies in the cluster, which we will 
call satellites.  For this reason, models of BCG formation routinely 
assume that BCGs are special (e.g. Milosavljevi\'c, \etal\ 2006;
De Lucia \& Blaizot 2007).
And there is growing evidence from scaling relations that BCGs are 
indeed special:  they have larger sizes, smaller velocity dispersions, 
smaller color gradients, and are less spherical than expected for 
their luminosities (e.g. Bernardi \etal\ 2011).  But whether or not 
this has left a distinct signature in the distribution of BCG 
luminosities is an open question.  

There are at least two reasons why the answer is interesting.  
First, Halo Model (see Cooray \& Sheth 2002 for a review) interpretations 
of the luminosity dependence of galaxy clustering explicitly distinguish 
between the central and the other (satellite) galaxies in a halo.  
If BCGs are just the brightest of a universal galaxy luminosity 
function, then including this constraint significantly simplifies 
such analyses.  

The second is that the distribution of BCG luminosities is considerably 
narrower than that of all galaxies, so the large number of clusters that 
will soon be available may permit BCGs to provide a useful consistency 
check of standard candle constraints on the luminosity-distance relation 
(Hubble 1936; Scott 1957; Sandage, Kristian \& Westphal 1976).  
Alternatively, if the $d_{\rm L}(z)$ relation is known and BCGs are just 
statistical extremes, then the BCG luminosity function can be used to 
provide a rather accurate determination of the evolution of the galaxy 
luminosity function.  Since BCGs are bright, they are amongst the 
easiest targets for spectroscopy, so this measure of galaxy evolution 
comes at a small fraction of the cost of a full galaxy survey.  
On the other hand, if they are special in other ways, one must also 
understand how the physical processes which affected their formation 
evolve.  

In what follows we will use the Halo Model analyses to motivate why 
one might have thought that the BCG luminosity function might simply 
be given by the extreme value statistics of the all galaxy luminosity 
function, even though centrals are often explicitly treated as being 
special.  
In this case, it is natural to ask if the luminosity function of the 
brightest satellite galaxies -- the BSG luminosity function -- is 
simply given by the associated order statistics of the second brightest 
(rather than the brightest) object.  
Then we will show that Halo Model analyses also suggest that the BSG 
luminosity function should be given by the extreme value statistics 
of a satellite galaxy luminosity function, rather than the order 
statistics of the all galaxy luminosity function.  
We will show that both statements cannot be right, so we turn to the 
data to decide the issue.

Section~\ref{prelims} summarizes the main results from Order
Statistics and the Halo Model approach which are relevant to this
study.
Section~\ref{sdss} uses a group catalog of SDSS galaxies 
from Berlind \etal\ (2006) (henceforth B06) 
to test these models regarding BCGs and BSGs.  
These tests include the luminosity functions of BCGs and BSGs, 
the ratio of the luminosities of the first and second brightest 
galaxies in a cluster, and the luminosity function of satellite 
galaxies (i.e., not just the BSGs).  
Section~\ref{marks} shows that luminosity weighted clustering 
provides a novel test of Order Statistics.  
A final section summarizes.  
Appendix~\ref{psatExstat} provides a derivation of the satellite 
distribution if the BCGs satisfy extreme value statisics.  
Appendix~\ref{shuffled} demonstrates that tests of order statistics 
are better made holding the number of galaxies fixed, rather than other
parameters such as group luminosity, etc.  
And Appendix~\ref{zehaviHOD} describes the details of the Halo Model 
that are relevant to our study. 

\section{Preliminaries}\label{prelims}

\subsection{Extreme value and order statistics}\label{exvals}

The probability that the largest of $N$ independent draws from a 
distribution $p(L)$ is less than $L$ is given by 
\be
 g_1(<L|N) = P(<L)^N\,,
\label{g1cum}
\ee
where $P(<L) = \int_{L_{\rm min}}^L {\rm d}L^\prime p(L^\prime)$, and 
we will define the value of $L_{\rm min}$ later.  
The associated differential distribution is  
\be
 g_1(L|N) = N\,p(L)\, P(<L)^{N-1}\,.
\label{g1gal}
\ee
Similarly, the distribution of the second largest of the $N$ draws 
satisfies 
\be
 g_2(<L|N) = P(<L)^{N} + {N\choose 1}\,P(>L)\,P(<L)^{N-1}\,,
\ee
so that 
\be
 g_2(L|N) = N(N-1)\,p(L)\,P(>L)\,P(<L)^{N-2}\,.
\label{g2gal}
\ee
More generally, the $n$th largest of $N$ draws obeys 
\be
 g_n(L|N) = {N\choose n}\,np(L)\,P(>L)^{n-1}\,P(<L)^{N-n}\,.
\label{gngal}
\ee
In what follows we will be interested in whether or not the 
distribution of BCG and BSG luminosities are given by 
equations~(\ref{g1gal}) and~(\ref{g2gal}) respectively, with 
$p(L)$ given by the luminosity function of all galaxies, 
although clearly, similar tests with the other values of $n$ could 
also be devised.

If we define the satellite luminosity function as the distribution 
which is obtained by subtracting the distribution of centrals 
(assumed to be $g_1(>L|N)$) from that of all galaxies, then one might 
ask if the luminosity function of BSGs is given by inserting this 
distribution of satellite luminosities for $p$ in the expression 
above for $g_1$.  In Appendix~\ref{psatExstat}, we show that this 
is not the same as $g_2$.  We will argue shortly that the Halo Model 
suggests that both of these models might be acceptable for satellites.

When discussing the group catalog of B06, we will 
use the luminosity distribution \pgal\ derived from equation~(9) of 
Bernardi \etal\ (2010), restricted and normalized to $L>L_{\rm  min}$,
for some $L_{\rm min}$ which we will specify shortly. Explicitly, 
\be
L\,\pgal = \frac{\beta\,(L/L_{\ast})^{\alpha}\,{\rm e}^{-(L/L_\ast)^\beta}}
  {\Gamma\left(\alpha/\beta,(L_{\rm min}/L_{\ast})^\beta\right)} .
\label{pgal-bernardi}
\ee
The parameters in Table B1 of Bernardi \etal\ assumed that absolute
magnitudes brighten as $1.3z$, and were quoted at $z=0$. Since the B06
sample is based on absolute magnitudes that were $k$-corrected and
evolution corrected to $z=0.1$, we adopt an increased value $L_{\ast}$
(by a factor $10^{0.4\times1.3\times0.1}$) but leave the other
parameters $\alpha$ and $\beta$ unchanged. I.e., we use,
\be
L_{\ast}=0.104 \times10^{10}\Lsun h^{-2} ~;~ \alpha=1.12 ~;~
\beta=0.533 \,.
\label{params-bernardi}
\ee
This fit differs slightly from the usual Schechter function of Blanton
\etal\ (2003) (their Table 2): 
\be
L_{\ast}= 1.202\times10^{10}\Lsun h^{-2} ~;~ \alpha=-1.05 ~;~ 
\beta=1\,.
\label{params-blanton}
\ee

Before moving on, note that the median value of $g_1$ occurs at 
luminosity $L_{1/2}$, where 
\be
 (1/2)^{1/N} = P(<L_{1/2}).
\ee
If $p(L)$ were simply an exponential distribution, then 
$P(>L)=\exp[-(L-L_{\rm min})/L_\ast]$, so 
$-\ln(2)/N = \ln[1 - P(>L_{1/2})] \approx - P(>L_{1/2})$, 
and 
 $L_{1/2}/L_\ast = L_{\rm min}/L_\ast - \ln\ln(2) + \ln(N)$ 
would grow logarithmically with $N$.  
It is the fact that $\ln L_{1/2}$ is an even weaker function of $N$ 
which makes BCGs standard candles (if they are indeed just statistical 
extremes). 

Similarly, if we define $L_{0.84}$ by 
 $0.84 = P(<L_{0.84})^N$,
then the difference between it and $L_{1/2}$ is a measure of the width 
of $g_1$.  For an exponential distribution this is given by 
 $\Delta L = L_{0.84}-L_{1/2} = L_\ast\ln[(1-(0.5)^{1/N})/(1-(0.84)^{1/N})]
           \approx 1.38\,L_\ast$. 
If $L_{\rm min}/L_\ast \ll 1$, this means that $\Delta L/L_{1/2}$ 
decreases as $1.38/\ln(N)$, meaning BCGs in richer clusters are better 
standard candles.  

If $p$ is given by equation~(\ref{pgal-bernardi}) using
\eqn{params-bernardi} and $L_{\rm min}=0.747 \times 10^{10}\Lsun
h^{-2}$, then $L_{1/2} = (3.24,3.71)\times 10^{10}\Lsun h^{-2}$ and  
 $\Delta L = (1.69,1.78) \times 10^{10}\Lsun h^{-2}$ for $N=(10,15)$, 
respectively (see also \fig{fig-shuffled}).  

Notice that order statistics such as these are explicitly a function 
of the number of draws $N$.  Therefore, the natural way to test if, 
e.g., BCGs are well-fit by equation~(\ref{g1gal}) with $p(L)$ given 
by the all-galaxy luminosity function of equation~(\ref{pgal-bernardi}) 
is to perform tests on groups which have the same $N$ (rather than 
the same mass, total luminosity, etc.).  We discuss this further in 
Appendix~\ref{shuffled}.

\subsection{The Halo Model approach}\label{halomodel}
The Halo Model approach assumes that galaxies reside in dark matter 
halos, each of which may host more than one galaxy.  Measurements of 
how the number density and clustering strength depend on galaxy 
luminosity are used to determine how galaxies populate halos (e.g., 
$\Ngal\propto M_{\rm halo}$?)  There are two ways in which this is 
usually done.  

The first is known as the Halo Occupation Distribution (HOD) approach 
(e.g. Zehavi \etal\ 2005, 2011).  Here, one determines how the number of 
galaxies brighter than some $L_{\rm min}$ must scale with halo mass so 
as to produce the observed number density and clustering strength.  
In detail, the fitting assumes that, in halos which host at least 
one galaxy, the distribution of the number of additional galaxies 
follows a Poisson distribution.  Since a Poisson distribution has 
one free parameter, the HOD approach aims to characterize how this 
parameter depends on halo mass and $L_{\rm min}$.  
By repeating the analysis for a range of $L_{\rm min}$ values, this 
approach, in effect, provides a determination of how the galaxy 
luminosity function depends on halo mass.  The additional assumption 
that a halo must contain a central galaxy before it can host 
satellites allows one to interpret the HOD findings in terms of 
centrals and satellites.  In what follows we will describe a few 
conclusions which follow from this assumption. The ease with which 
such conclusions can be derived has meant that it is now conventional 
to even phrase the HOD explicitly in terms of a 
`central + Poisson-satellites' model, before fitting to data. We discuss
a simplified HOD model below to emphasize our main points; in Appendix
C we describe the more sophisticated implementation from Zehavi et 
al. (2011), which we will use for comparison with data.

The HOD analysis assumes that the number of satellites brighter than 
$L$ scales as a power-law in halo mass, 
 $[M/M_1(L)]^{\alpha(L)}$, 
where $M_1$ is the halo mass scale on which each halo hosts, on average, 
one satellite.  
Therefore, the distribution of satellite luminosities brighter than 
some $L_{\rm min}$, that are in halos of mass $M$, is simply 
\be
P^{\rm (HOD)}_{\rm sat}(>L|M,L_{\rm min})= 
  \frac{[M/M_1(L)]^{\alpha(L)}}{[M/M_1(L_{\rm min})]^{\alpha(L_{\rm  min})}}.
\label{PsatHOD}
\ee
Matching the observed luminosity dependence of clustering requires that 
\be
 \frac{M_1(L)}{10^{12} h^{-1}M_\odot}\approx 23\,M_{\rm min}(L) 
      = 23\,[\exp(L_{10}) - 1]\,,
\label{M1L}
\ee
where $L_{10}$ is the luminosity in units of $10^{10}h^{-2}L_\odot$, 
and $\alpha\approx 1$ approximately independent of $L$ (it increases 
by about 20 percent when $L_{\rm min}$ increases by 2 orders of magnitude).   
The fact that $\alpha$ is approximately independent of $L$ implies 
that the distribution of satellite $L$ must be approximately 
independent of halo mass $M$, and this is in good agreement with 
measurements in the SDSS (Skibba \etal\ 2007).

The inverse of the $M_{\rm min}(L)$ relation (equation~\ref{M1L}) 
indicates how the luminosity of the central galaxy scales with halo 
mass:
\be
 \frac{L_{\rm cen}}{1.1\times 10^{10}h^{-2}L_\odot} \approx 
 \ln\left(1 + \frac{M}{10^{12} h^{-1}M_\odot}\right).
\label{LcenM}
\ee
Note that although $L_{\rm cen}$ is a strong function of mass at small 
$M$, it grows only logarithmically at high masses.  This will be important 
shortly.

There is another implementation of the Halo Model which is known as 
the Conditional Luminosity Function (CLF) approach 
(e.g. Yang, Mo \& van den Bosch 2008).  
In this case, one models how the galaxy luminosity function depends on 
halo mass by explicitly postulating that the centrals and satellites 
have different (halo mass dependent) luminosity functions.  
The HOD and CLF analyses yield consistent conclusions.  E.g., Table~2 
of Yang \etal\ (2008) shows that their analysis of how the mean number of 
satellites scales with halo mass also yields $\alpha\approx 1$, and 
depends only weakly on halo mass.  Their $M_1(L)$ relation is reasonably 
well described by our equation~(\ref{M1L}).  And they find that the 
luminosity function of centrals becomes a weak function of halo mass at 
large masses (see their Figure~6).  Although they parametrize this using 
a double power-law, we have checked that the HOD-based expression above 
(our equation~\ref{LcenM}) provides a reasonable description of their 
fit (their equation~6).  

\subsection{Extreme values and the Halo Model}\label{exhalos}
The weak, approximately logarithmic growth of $L_{\rm cen}$ at large  
halo mass suggests that BCGs might be drawn from an extreme value 
distribution that is determined from the full galaxy distribution 
(equation~\ref{g1gal} with~\ref{pgal-bernardi}).  
In this case, it is natural to ask if the luminosity distribution 
of BSGs is given by equation~(\ref{g2gal}) with~(\ref{pgal-bernardi}).  

On the other hand, if the distribution of satellite luminosities has 
an approximately universal form (i.e. halo mass determines the overall 
number but not the shape of the distribution), then one might 
reasonably expect extreme value statistics to provide a good 
description of BSG luminosities.  
In particular, one would therefore expect that the luminosity 
distribution of the \emph{second} brightest galaxy in a group, which 
is the brightest satellite, should be well-described using the extreme 
values of \psat, where \psat\ is the universal satellite luminosity 
function (rather than that for all galaxies):  e.g., if the HOD parameter 
$\alpha$ were indeed independent of $L$, then the BSG distribution 
should be given by using equation~(\ref{PsatHOD}) in equation~(\ref{g1gal}).  
However, as we remarked earlier (see Appendix~\ref{psatExstat} for details),  
this is, in general, different from the expected distribution of the 
\emph{second} brightest galaxy in a group, under the assumption that 
the BCGs are just the brightest drawn from the universal galaxy 
luminosity function. 

\section{Comparison with SDSS data sets}\label{sdss}
To address these issues, we use the Mr20 catalog from 
Berlind \etal\ (2006), which is based on the SDSS main galaxy catalog.  
The groups in the Mr20 catalog are derived from a volume limited
sample containing galaxies with spectroscopic redshifts between
$0.015<z<0.1$, with an $r$-band absolute magnitude ($k$-corrected and
evolution corrected to $z=0.1$) brighter than
$M_{0.1r}-5\log_{10}h\approx-19.9$\footnote{The evolution correction
  brightens the magnitude of the $j^{\rm th}$ object according to
  $M_{0.1r,j}\to M_{0.1r,j}+Q(z_j-0.1)$, where $Q=1.62$ is taken from
  Blanton \etal\ (2003, their Figure 5). Note that the values reported
  in the public catalog are not evolution corrected; all our results
  use the corrected absolute magnitudes. Also, we find that the
  value of the absolute magnitude threshold in the public catalog
  after evolution correction is $-19.924$, which is what we adopt.},
corresponding to luminosity $L_{\rm min}=0.747 \times 10^{10}\Lsun
h^{-2}$. The groups are identified by the Friends-of-Friends
algorithm, corrected for fiber collisions and survey-edge effects (see
B06 for details), which results in $4107$ groups containing $\Ngal>2$
members, i.e. $\Nsat>1$ satellites. As noted by B06, however, the
smaller groups with $\Ngal<10$ are affected by systematic biases in
the identification algorithm. We therefore restrict our extreme values
analysis to the subset of the Mr20 catalog containing groups with
$\Ngal\geq10$ members, of which there are $332$. The catalog contains
luminosity information for all member galaxies.

\subsection{Group-galaxy luminosity distribution}\label{lumgals}
\begin{figure}
 \centering
 \includegraphics[width=0.95\hsize]{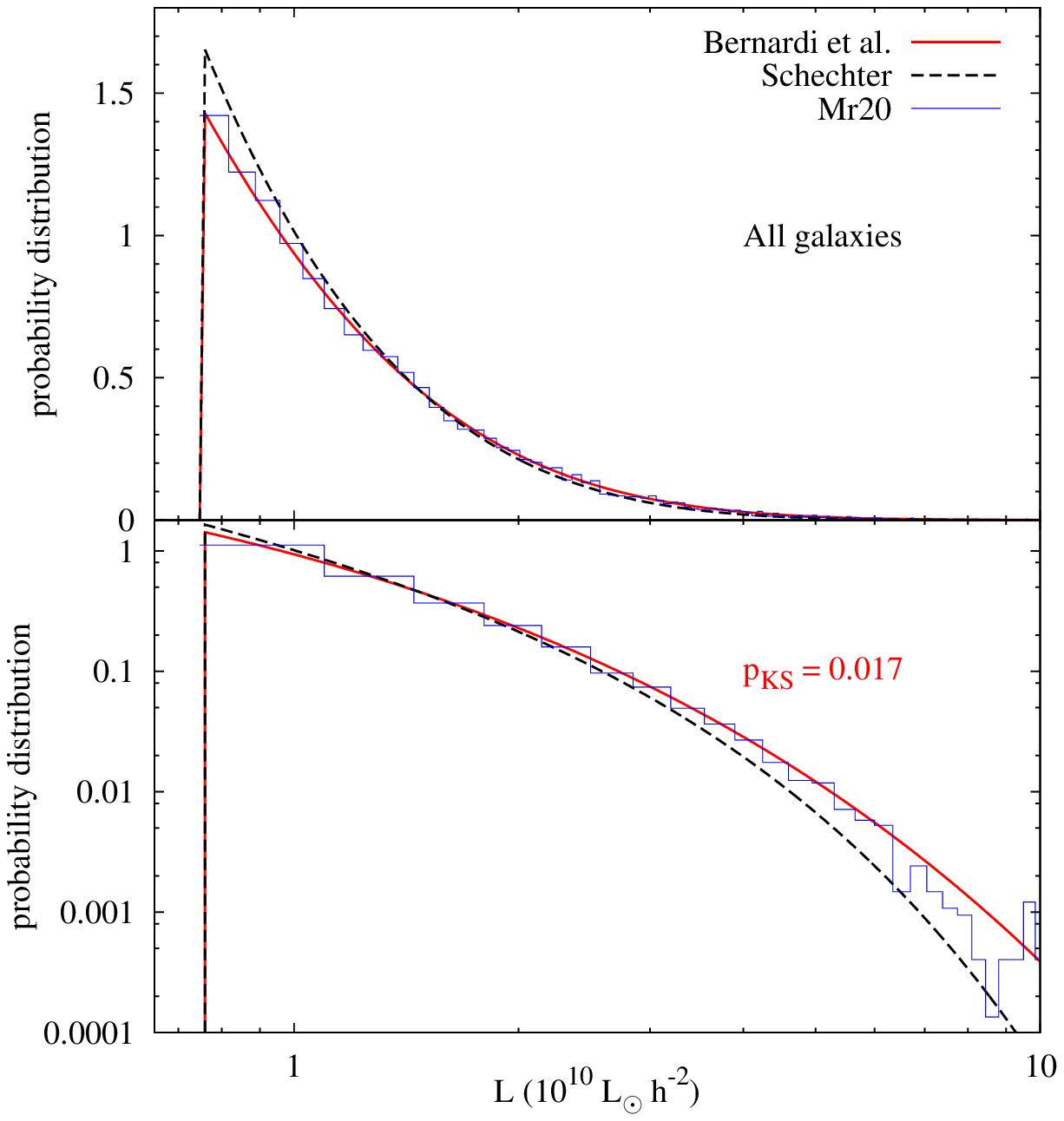}
 \caption{Luminosity distribution of the Mr20 galaxy catalog from
   Berlind \etal\ (2006) which contains groups with $\Ngal>2$ members
   (histogram). Solid curve shows \eqn{pgal-bernardi}) with parameters 
   from \eqn{params-bernardi}. We also display the $p$-value from a KS
   analysis comparing the data with the solid curve. 
   For comparison, dashed curve, which 
   predicts a slightly fainter distribution than observed, 
   shows the Schechter distribution with parameters from 
   Blanton \etal\ (2003). }
 \label{fig-pgal}
\end{figure}

We begin by asking whether the galaxies in the Mr20 catalog are
consistent with being drawn from a universal luminosity distribution. 
\fig{fig-pgal} shows the galaxy luminosity distribution of the full 
Mr20 catalog, which has groups containing $\Ngal>2$ members. 
The solid curve shows equation~(\ref{pgal-bernardi}) with parameters 
adapted from Bernardi \etal\ (2010) (equation~\ref{params-bernardi}).
It describes the luminosity distribution in the Mr20 sample remarkably 
accurately, especially given that this fit was obtained from the full
galaxy sample -- not the subset which satisfy the B06 selection
criterion. We also display the $p$-value from a Kolmogorov-Smirnov
(KS) analysis comparing the solid curve with the data. 

For comparison, the dashed curve shows equation~(\ref{pgal-bernardi}) 
with parameters derived from Blanton \etal\ (2003)
(equation~\ref{params-blanton}). 
This form
predicts somewhat fainter galaxies than are present in the B06 sample.  
This is not surprising given previous work showing that that the 
Schechter parametrization is a poor description of the bright-end 
of the all-galaxy sample, and groups are expected to preferentially 
host the brightest galaxies. (A KS analysis in this case gives a
negligibly small $p$-value.) 

\begin{figure}
 \centering
 \includegraphics[width=\hsize]{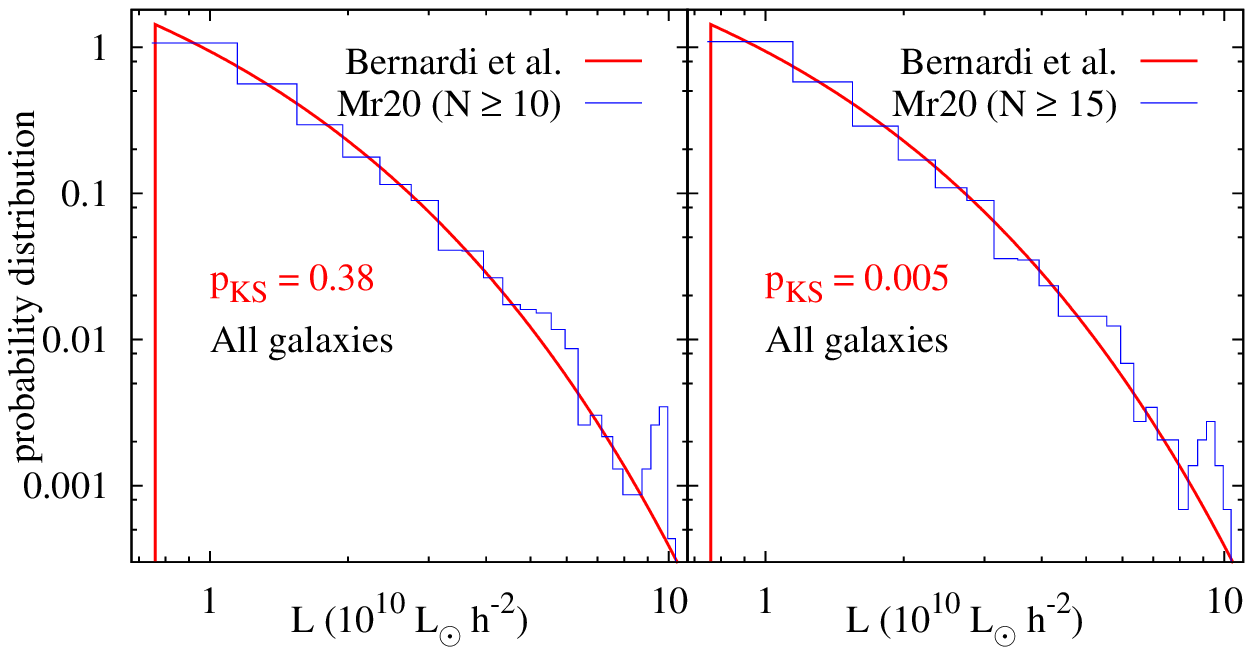}
 \caption{Luminosity distribution of subsamples of the Mr20 galaxy
   catalog containing groups with $\Ngal\geq10$  and $\Ngal\geq15$
   members (histograms). Solid curve in each 
   panel is the same as in \fig{fig-pgal}, and shows 
   \pgal\ (equation~\ref{pgal-bernardi} using
   equation~\ref{params-bernardi}). This same function provides an
   accurate description of each subsample. We also display the
   $p$-values from a KS analysis comparing the data with the solid
   curve.} 
 \label{fig-pgal-sub}
\end{figure}

As mentioned previously, we do not expect to obtain robust results for
groups with $\Ngal<10$ and we will henceforth discard these objects.
\fig{fig-pgal-sub} shows the luminosity distributions in two
subsamples constructed by restricting to $\Ngal\geq10$ and
$\Ngal\geq15$ members. Equation~\eqref{pgal-bernardi} with parameters
from \eqn{params-bernardi} is an accurate description of each 
subsample. We show the KS $p$-values from a comparison of each 
subsample with this distribution. This suggests that these B06 
group galaxies are in fact drawn from a universal luminosity 
distribution which is approximately independent of group richness 
and halo mass.  

It is not a priori obvious that this distribution should have been 
universal, nor is it obvious that this universal function should have 
simply been the all-galaxy luminosity function (truncated to only 
include objects that are brighter than $M_r<-19.9$).  The latter is 
worrying, because it might arise as follows.  Separating true group 
members from interlopers on the basis of angular position and redshift 
space information alone is known to be difficult.  Using colors to 
help identify members of the same group helps significantly, but the 
B06 algorithm does not use color information.  
So it is conceivable that the algorithm returns an accurate estimate 
of the number of members in a group, without actually identifying the 
members themselves correctly.  In this case, it is possible that 
objects identified in the catalog as group members are effectively 
random samples of the underlying galaxy distribution -- with the 
choice of link-length parameters having been (too) strongly influenced 
by theoretical expectations about what the distribution of $N$ should 
be.  If this is indeed why the all-galaxy luminosity function describes 
the B06 luminosities, then it would invalidate a number of published 
analyses based on this catalog -- particularly those which seek to 
constrain the phenomenon known as assembly bias.  
Section~\ref{marks} describes a novel test of this possibility, which 
uses mark correlations.  

In the remainder of the paper, when discussing the Mr20 catalog, we
will show results obtained assuming universality of \pgal\ using
\eqn{pgal-bernardi} with parameters from \eqn{params-bernardi}.

\subsection{Central galaxies as statistical extremes}
\label{exvalgals}

\begin{figure}
 \centering
 \includegraphics[width=\hsize]{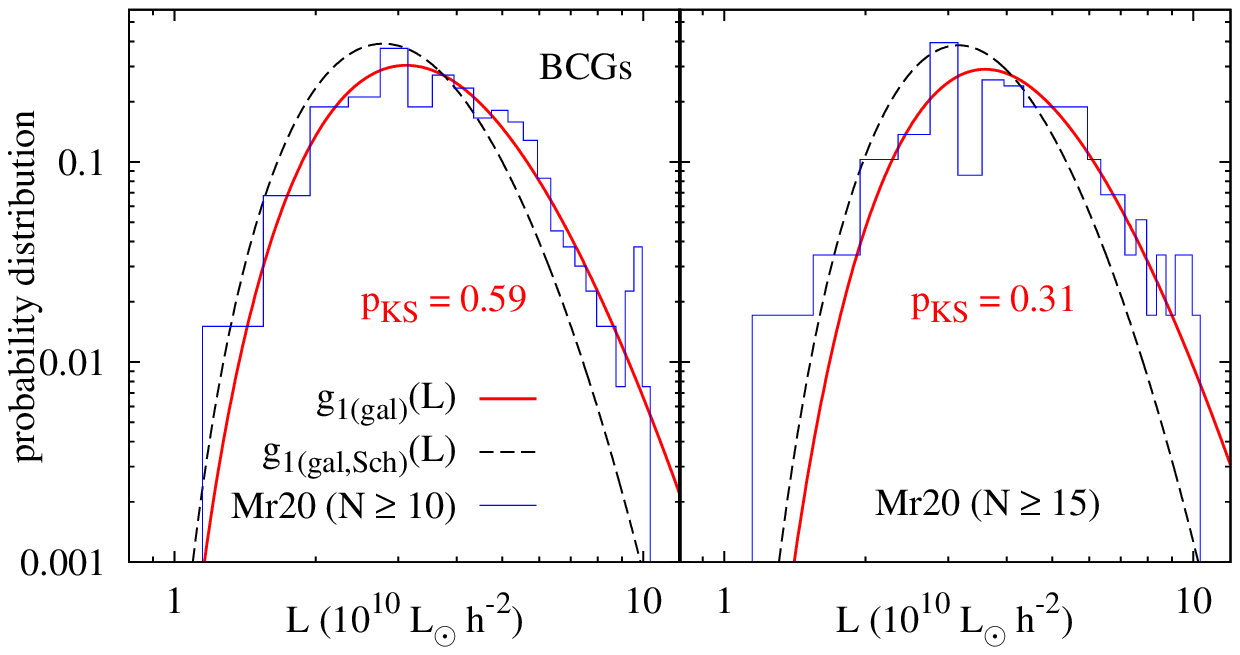}
 \caption{Luminosity distribution of the BCGs in subsamples of the
   Mr20 catalog with $\Ngal\geq10$ and $\Ngal\geq15$
   (histograms). Solid curves show the extreme value statistics for the
   brightest object in \Ngal\ independent draws from the distribution
   \pgal\ of \eqn{pgal-bernardi} with parameters from
   \eqn{params-bernardi}, averaged over the observed distribution of 
   \Ngal\ in each subsample as in \eqn{g1galavg}. We also display the
   corresponding $p$-values from a KS analysis comparing the data with
   the solid curves.  For comparison,
   dashed curves show the corresponding predictions using the
   Blanton et al. parameters from \eqn{params-blanton}.} 
 \label{fig-g1galavg-Mr20}
\end{figure}%

We now ask if the BCG of a group with \Ngal\ members can be described  
as the brightest of \Ngal\ independent draws from \pgal.
Since the number of groups in the Mr20 catalog with exactly
\Ngal\ galaxies where $\Ngal\geq10$ is small, we make this comparison 
by constructing subsamples restricted to a minimum value
of $\Ngal$ as in section \ref{lumgals} above. In this case, the 
corresponding theory prediction must be averaged over the allowed 
values of \Ngal. We do this by constructing the distribution 
$p(\Ngal)$ directly from the given subsample, and averaging
\eqn{g1gal} over this distribution: 
\be
 g_1^{\rm (gal)}(L) = \sum_{{\rm allowed}\ \Ngal} g_1^{\rm (gal)}(L|\Ngal)\,p(\Ngal)\,.
\label{g1galavg}
\ee
\fig{fig-g1galavg-Mr20} shows the results for subsamples of Mr20
containing $\Ngal\geq10$ and $\Ngal\geq15$ members.  The solid red
curves show the extreme value prediction \eqref{g1galavg} using
\eqn{pgal-bernardi} with parameters from \eqn{params-bernardi},
together with the respective $p$-values from a KS analysis. We see
a remarkable agreement between the predictions
and the data. (Of course, $p(\Ngal)$ is different for each subsample.)
For comparison we also show the predictions when using the Blanton 
\etal\ parameters from \eqn{params-blanton}. (In this case the KS
analysis gives tiny $p$-values $\lesssim10^{-7}$.)

\subsection{Satellite luminosity distribution}\label{lumsats}
We now analyse the satellites in the Mr20 catalog, i.e. the
data set obtained by subtracting the BCGs from the full catalog. 
The results of the previous section allow us to use the satellite
luminosity distribution to perform a consistency check. If the full
galaxy luminosity distribution \pgal\ is indeed universal, \emph{and}
if the BCGs obey the extreme value statistics of \pgal, then the 
satellite luminosity distribution of the subsamples with $\Ngal\geq10$
and $\Ngal\geq15$ must depend on the minimum number of members
in a specific way: the predicted distribution is the average of
\eqn{app-psat}, weighted by the number of satellites $\Ngal-1$, over
the observed distribution of $\Ngal$. The dotted curves show these
predictions; they provide a good description of the data. The
associated $p$-values from a KS test are also shown. 

\begin{figure}
 \centering
 \includegraphics[width=\hsize]{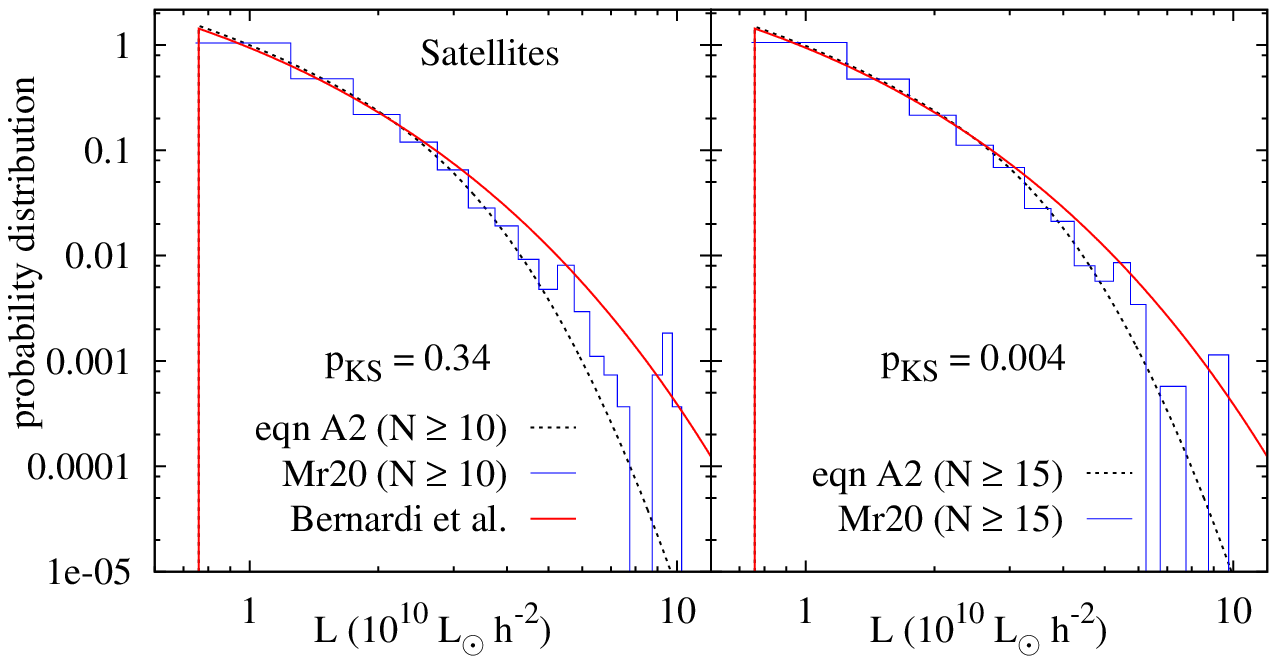}
 \caption{Luminosity distribution of the satellites in subsamples of
   the Mr20 catalog with $\Ngal\geq10$ and $\Ngal\geq15$
   (histograms). 
   Dotted curves show the result of assuming that the 
   satellite distribution is given by subtracting from the full 
   galaxy distribution the associated extreme value distribution, 
   averaged over the observed distribution of $\Ngal$ 
   (the average of equation~\ref{app-psat}, weighted by the number of
   satellites $\Ngal-1$, over $\Ngal$). The KS $p$-values associated
   with the dotted curves are also displayed.
   Solid curve in each panel is the full galaxy distribution
   \pgal\ of \eqn{pgal-bernardi} with \eqn{params-bernardi}. 
   }
 \label{fig-psat}
\end{figure}

The solid curves (same in each panel) show \pgal\ from 
\eqn{pgal-bernardi} with \eqn{params-bernardi}. The analysis in
Appendix~\ref{psatExstat} predicts that the satellite distributions at
large \Ngal\ should become approximately universal, asymptotically
approaching the distribution \pgal, and we see that this is indeed the
case. 

\subsection{Brightest satellites as second brightest objects}
\label{exvalsats}
The brightest satellites are, in a sense, more interesting than 
the BCGs, because there are \emph{two} statistical contenders for
describing their luminosity distribution. 

\begin{figure}
 \centering
 \includegraphics[width=0.95\hsize]{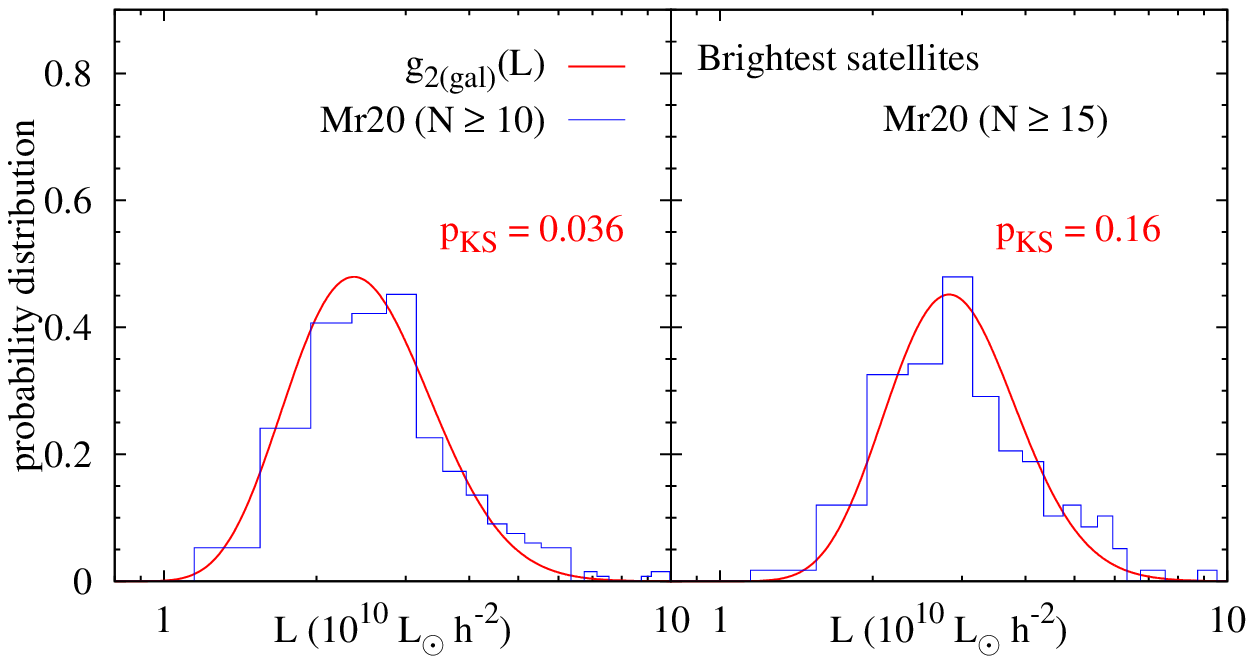}
 \caption{Luminosity distribution of the brightest satellites in 
   subsamples of the Mr20 catalog containing $\Ngal\geq10$ and
   $\Ngal\geq15$ members (histograms). 
   Solid curves show the order statistics for
   the \emph{second} brightest object in \Ngal\ independent draws from
   the distribution \pgal\ of \eqn{pgal-bernardi} with
   \eqn{params-bernardi}, averaged over the observed distribution of
   \Ngal, similarly to \eqn{g1galavg} but using the extreme value
   distribution from \eqn{g2gal}. The corresponding KS $p$-values are
   also displayed.} 
 \label{fig-g2galavg-Mr20}
\end{figure}

The first is motivated by the fact that BCGs appear to be consistent 
with extreme value statistics (equation~\ref{pgal-bernardi} in 
equation~\ref{g2gal}).  This suggests that the BSGs ought to be well 
described as the \emph{second} brightest of \Ngal\ independent draws 
of the universal galaxy distribution \pgal\ (equation~\ref{pgal-bernardi} 
with equation~\ref{params-bernardi} in equation~\ref{g2gal}).  
\fig{fig-g2galavg-Mr20} shows that this is indeed the case.

\begin{figure}
 \centering
 \includegraphics[width=0.95\hsize]{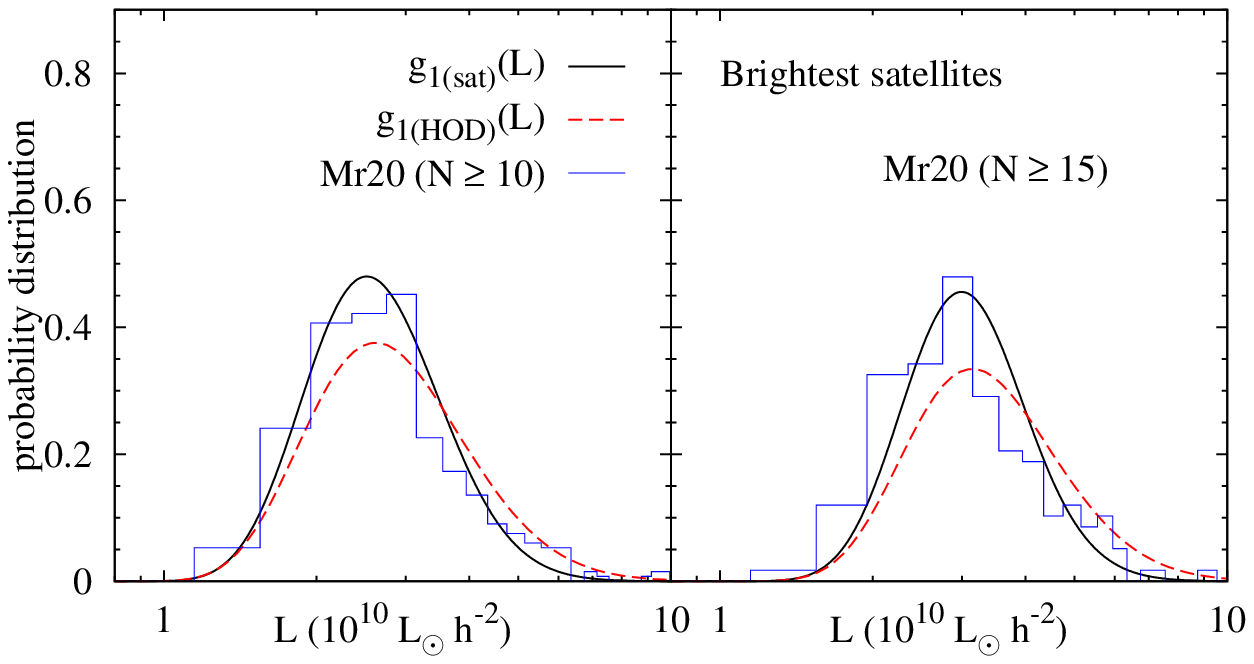}
 \caption{Luminosity distributions of the brightest satellites in two
   subsamples of the Mr20 catalog: histograms are the same as
   in \fig{fig-g2galavg-Mr20}. Solid black curves show the extreme
   value statistics for the brightest object in \Nsat\ independent draws
   from the distribution \psat.  We used extreme value predictions of
   the corresponding dotted curves of \fig{fig-psat} (i.e., $g_{\rm
     1sat}(L|N)$ from equation~\ref{g1sat}, averaged over the observed
   $\Nsat$ values). 
   Dashed red curves show the analogous extreme value distributions
   based on the HOD model of Zehavi \etal\ (2011) described in
   Appendix~\ref{zehaviHOD}.} 
 \label{fig-g1satavg-Mr20}
\end{figure}

\subsection{Brightest satellites as statistical extremes}
The alternative model is that BSGs are extremes of some universal 
satellite distribution, as suggested by the HOD model, and by the 
measurements which suggest that the overall satellite luminosity 
function is basically independent of group richness 
(Skibba \etal\ 2007; Hansen \etal\ 2009).
\fig{fig-g1satavg-Mr20} shows this comparison for two choices of the
universal distribution \psat.  More precisely, we compare the
brightest satellite distributions for the two subsamples (the
histograms are the same as in \fig{fig-g2galavg-Mr20}), with the
extreme value distribution appropriate for the brightest of
\Nsat\ independent draws of an assumed universal satellite
distribution \psat, averaged over the observed distribution of
\Nsat\ values (solid and dashed curves, see below).   
The resulting extreme value distribution is identical in form 
to \eqns{g1galavg} and \eqref{g1gal}, with $\Ngal\to\Nsat$ and 
$\pgal\to\psat$.  These predictions depend on \psat.  

The solid curves show the result of using our analytic calculation of
\psat\ (dotted curves in \fig{fig-psat}) to provide a
calculation of $g_{\rm 1sat}(L|N)$ (from equation~\ref{g1sat}) which
we then average over the observed distribution of \Nsat. 
The solid curves provide a good description of the 
measurements, with KS $p$-values comparable to those in \fig{fig-g2galavg-Mr20}; this is not surprising since we are at large \Ngal, where we know that $g_{\rm 1sat}\to g_{\rm 2gal}$ (see discussion at 
the end of Appendix~\ref{psatExstat}). Unfortunately then, this large \Ngal\ dataset is not suitable for distinguishing between these two predictions.

The dashed curves show the result of using the HOD result \eqn{hod-g1avg}. We see that this
tends to always predict brighter distributions than the 
ones observed (the corresponding KS $p$-values are $\sim10^{-5}$). 
However, since the HOD prediction relies on parameter values derived
from statistical fits (Zehavi \etal\ 2011), we would caution against
taking these results at face value. A more complete treatment would involve accounting for parameter errors in the HOD fit, but this is beyond the scope of this work.  

\begin{figure}
 \centering
 \includegraphics[width=\hsize]{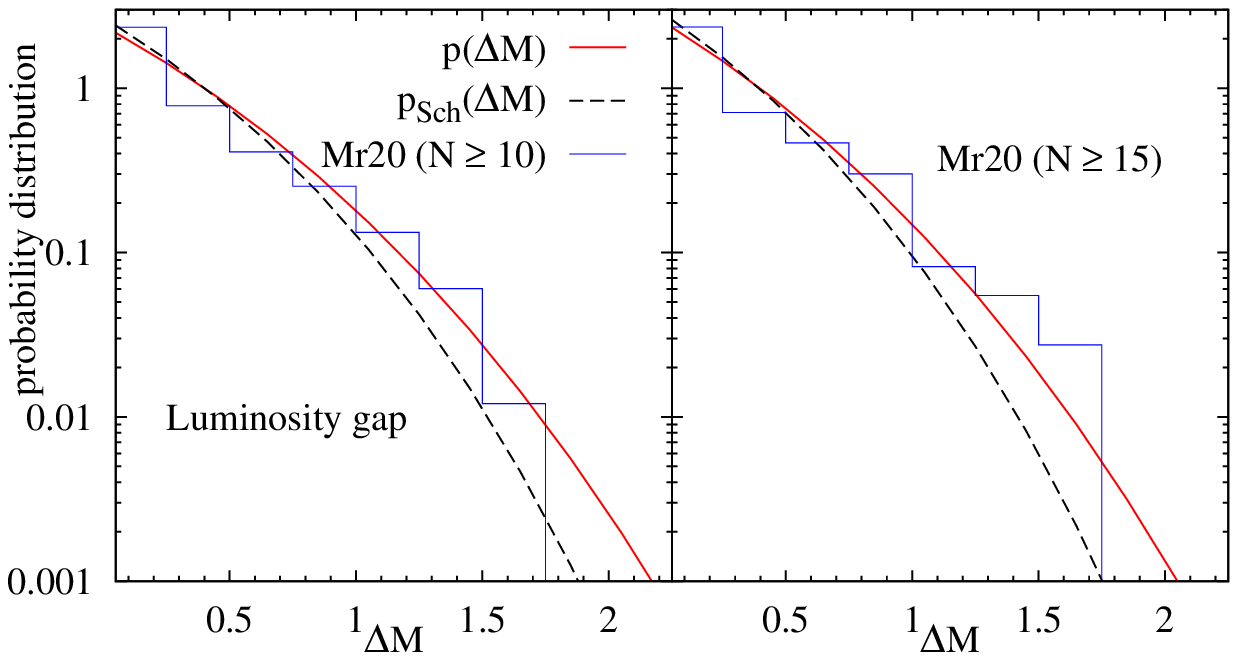}
 \caption{Distribution of the luminosity gap \DM\ for subsamples of
   the Mr20 catalog with $\Ngal\geq10$ and $\Ngal\geq15$
   (histograms). The solid curves are the corresponding predictions of
   extreme value statistics \eqn{pgapavg}, averaged over the
   corresponding observed distributions of \Ngal\ and using
   \pgal\ from \eqn{pgal-bernardi} with \eqn{params-bernardi}. For
   comparison, we also show the predictions based on parameters from
   \eqn{params-blanton}.} 
 \label{fig-gapavg}
\end{figure}

\subsection{The luminosity gap}
\label{gap}
If both the BCG and BSG luminosity functions are given by Order 
Statistics, then the joint distribution of their magnitudes
$p(M_1,M_2|\Ngal)$ (with $M_j$ being the $j^{\rm th}$ brightest) is 
\begin{align}
p(M_1,M_2|\Ngal) &= \Ngal(\Ngal-1)p_{\rm gal}(M_1)\nonumber\\
&\ph{\Ngal}\times p_{\rm gal}(M_2)
P_{\rm gal}(>M_2)^{\Ngal-2} \Theta(M_2-M_1)\,,
\label{M1M2joint}
\end{align}
with $\Theta(x)$ the Heaviside distribution, and we have used the same
notation $p_{\rm gal}$ as before to denote the universal galaxy
luminosity distribution, with $p_{\rm gal}(m) = L(m)p_{\rm
  gal}(L(m))\ln(10)/2.5$.  This allows us to derive the distribution
of the luminosity gap, defined as the difference in magnitudes
$\DM\equiv M_2-M_1$ of the second brightest and brightest
objects. Before considering this distribution, we note that another 
interesting distribution is that of the second brightest $M_2$, for a
fixed value of $M_1$. This is given by
\begin{align}
p(M_2|M_1,\Ngal) &=\frac{p(M_1,M_2|\Ngal)}{p(M_1|\Ngal)} \nonumber\\
&= (\Ngal-1)p_{\rm gal}(M_2) \frac{P_{\rm gal}(>M_2)^{\Ngal-2}}{P_{\rm 
    gal}(>M_1)^{\Ngal-1}} \Theta(M_2-M_1) \,.
\label{M2givenM1}
\end{align}
As $M_1$ is made more negative, i.e. as the BCG is made more luminous,
$P_{\rm gal}(>M_1)\to1$, and this distribution simply
asymptotes to that of the brightest of $N-1$ draws from the universal
galaxy distribution, independent of $M_1$. In other words, the typical
gap \DM\ increases as the BCG is made more luminous, illustrating
what is known as the Bautz \& Morgan (1970) effect. We can therefore
understand the latter as a straightforward consequence of order
statistics. 

The distribution of the luminosity gap \DM\ for fixed $N$ is predicted
to be 
\begin{align}
p(\DM|\Ngal) &= \Ngal(\Ngal-1) \int
dm \,p_{\rm gal}(m) \nonumber\\ 
&\ph{\Ngal(-1)}\times P_{\rm gal}(>m)^{\Ngal-2}p_{\rm gal}(m-\DM)\,.
\label{pgapfix}
\end{align}
As with the earlier distributions,
we can define averages of the fixed-\Ngal\ gap distribution over
subsamples, as
\be
p(\DM) = \sum_{{\rm allowed }\Ngal}p(\DM|\Ngal)p(\Ngal)\,.
\label{pgapavg}
\ee
\fig{fig-gapavg} compares the averaged prediction in \eqn{pgapavg}
with the subsamples of Mr20 having $\Ngal\geq10$ and $\Ngal\geq15$. 
We see a good visual agreement between the data and the prediction
based on \eqn{pgal-bernardi} with \eqn{params-bernardi} (solid
curves), except at small \DM, where the data show a number of groups
with \DM\ precisely equal to zero. We have checked that all these
systems correspond to fiber collided BCGs, and the algorithm of
B06 then assigns the brightest satellite both the redshift and the
absolute magnitude of the BCG. The KS $p$-values comparing the solid
curves to the full data (i.e., including the cases with $\DM=0$) are
negligibly small. However, comparing the same curves after subtracting
the spike from the data leads to large $p$-values ($>0.1$).
In either case (with or without the spike), using \eqn{pgal-bernardi}
with \eqn{params-bernardi} provides a better description of the large
$\DM$ tail than using \eqn{params-blanton}. 

\section{Mark correlations as a novel test of order statistics}\label{marks}
The previous section showed that the distribution of group-galaxy 
luminosities was very like that of all galaxies (whether or not they 
are in groups).  This raised the question of whether or not the group 
catalog has correctly separated group members from the field.  
We will use mark correlations to address this and related questions.  

The mark correlation $WW(r)/DD(r)$ is a statistic associated with 
pairs of objects separated by $r$;
 $DD(r)$ is the number of pairs of separation $r$, and 
 $WW(r)$ is the result of weighting each galaxy $i$ by some weight 
 $W_i/\bar W$, (where $\bar W$ is the average weight of the sample) 
when performing the pair count.  In what follows, we will use the 
luminosity as a weight.  If there is no correlation between a galaxy's 
luminosity and its position then $WW/DD$ should be unity; 
departures from unity indicate that the luminosities and positions 
are correlated.  

\begin{figure}
 \centering
 \includegraphics[width=\hsize]{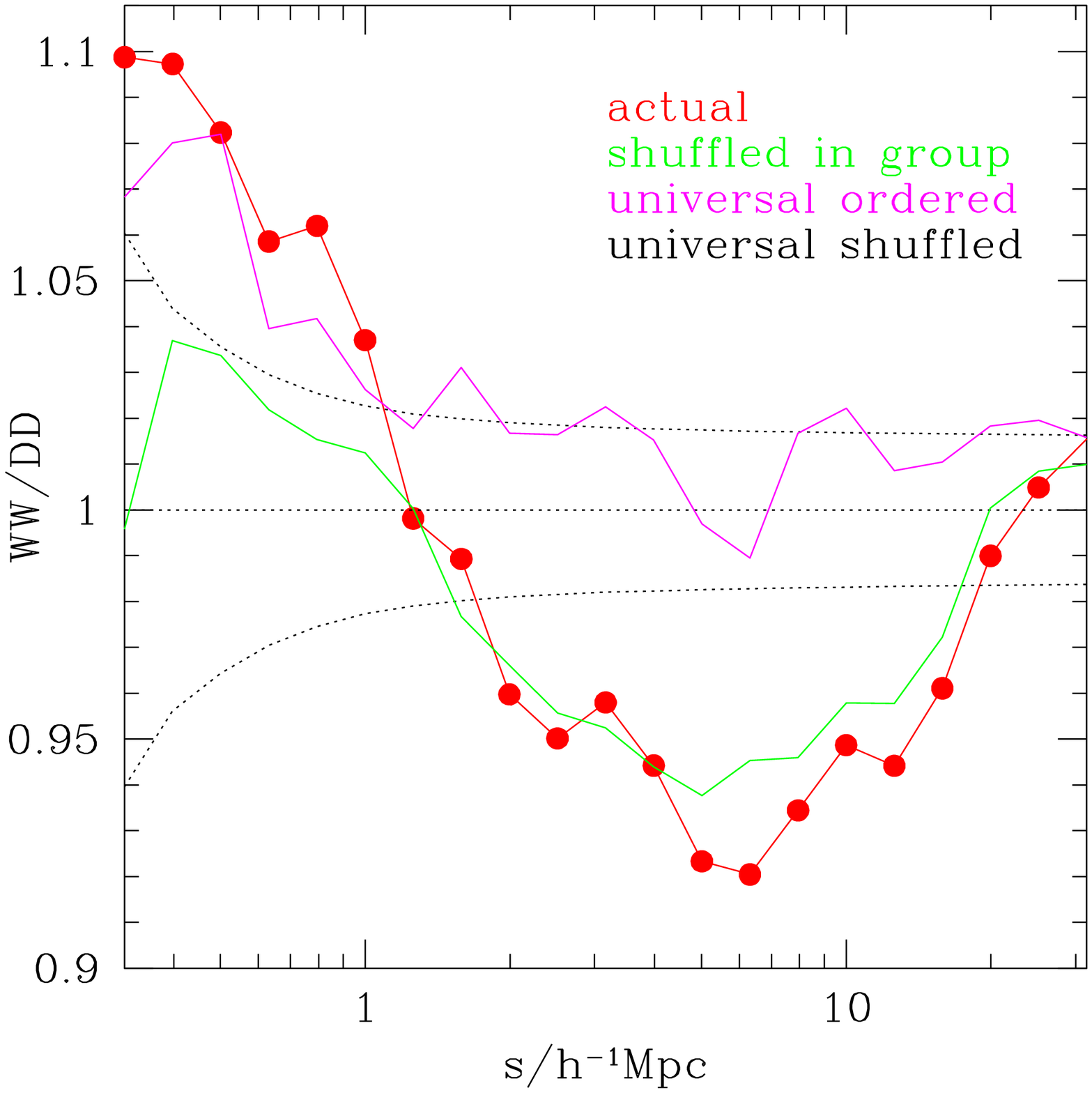}
 \caption{Luminosity mark correlation in the Mr20 catalog, restricted 
          to galaxies which reside in groups with more than 10 members
          (red symbols). The fact that this signal is different from unity on small scales ($<1h^{-1}$Mpc) indicates a 
          significant correlation between luminosity and location within the group. The fact that it is also different from 
          unity on larger scales shows that the luminosity function cannot be universal. This is more clearly demonstrated by 
          the `universal ordered' (magenta) and `shuffled in group' (green) curves which separate these two effects (see text). 
          The dotted curves (`universal shuffled') are a rough measure of the statistical significance of the difference from 
          unity.}
 \label{shuffledMarks}
\end{figure}%

We begin with the objects in the Mr20 group catalog.  If the 
group-identification process ended up assigning random galaxies to 
groups, then the catalog which results from scrambling the list 
of luminosities among all the members of the original catalog 
should be stastically the same as the original one.  Therefore, we 
measure the luminosity mark correlation $WW/DD$ in this scrambled 
catalog.  We find that $WW/DD\sim 1$ as it should.  
Repeating this scrambling procedure a few times and averaging the 
results gives some indication of the range of values around unity 
that one might expect given the observed set of $L$ and positions, 
if the luminosities and positions are indeed uncorrelated.  
The dotted curves in Figure~\ref{shuffledMarks} show this range, 
which we will use as a rough measure of the significance of a 
detection of a correlation between luminosity and position.  

Next, we scramble the luminosities as before -- thus erasing any 
possible correlations -- but then, within each group, we reassign the 
luminosities among the objects in the group so that they have the same 
rank-ordering as in the actual dataset.  The resulting measurement of 
$WW/DD$ shows the expected signal if the luminosity distribution were 
universal (i.e. the same for all groups), but the distribution within 
a group depends on location within the group (e.g., if the central 
galaxy is the brightest).  The figure shows the result of averaging
$WW/DD$ over many such `universal ordered' rescramblings (magenta curve). There is a clear tendency for 
$WW/DD > 1$ at $r\le 1h^{-1}$Mpc, indicating a stastically significant 
correlation between galaxy luminosity and group-centric position.  
Notice, however, that $WW/DD$ declines to $\sim 1$ on larger scales; 
this is a consequence of the fact that the typical group size is of 
order $r\le 1h^{-1}$Mpc, and we have erased all correlations on larger 
scales.  (In the Halo Model description of mark correlations (Sheth 2005), 
the two-halo term for this case is unity.)

Finally, the filled red symbols show $WW/DD$ in the original (unscrambled) 
group catalog.  On small scales, the signal is similar to that in the 
`universal ordered' groups, but it clearly differs from unity 
even on larger scales.  The fact that $WW/DD\ne 1$ on $>1h^{-1}$Mpc 
scales suggests that the luminosity function itself is not universal.  
To separate out the effects of the group-centric trends from possible 
group-group trends, we have performed a final scrambling.  
We keep the list of luminosities within a group the same, but, within 
each group, we scramble the luminosities amongst the group members.  
In this way, we remove that part of the signal from $WW/DD$ which is 
due to the correlation with group-centric position.  Notice that now 
$WW/DD$ (`shuffled in group', green curve) is like the original on large scales, but it does not increase 
as much on smaller ($<1h^{-1}$Mpc) scales.  

Thus, our mark correlation analysis has shown that 
 i) group-centric position matters; and 
 ii) even once this has been accounted for, the luminosity function is 
    not universal for all groups. 
The first point derives from the fact that the `universal ordered' 
signal is not unity on small scales, and it is very similar to the original one, and 
the `shuffled in group' signal differs from the original unscrambled 
one at these small scales.  This strongly suggests that the brightest cluster galaxy does 
indeed lie closer to (if not at) the group center.  
And the fact that, on larger scales, the actual signal differs so 
significantly from the `universal ordered' one (i.e. from unity) indicates that the 
luminosity function cannot be the same for all groups.  
This was not at all obvious from the (more traditional) analysis 
of the group and BCG luminosity functions in the previous section.

\section{Conclusions}
Brightest cluster galaxies are interesting for several reasons, the
simplest being that they are the brightest and therefore most easily
observable of galaxies. As mentioned in the Introduction, it is also
becoming increasingly certain that they are physically distinct in
several other ways, e.g. in their morphology, velocity dispersions,
etc., from the other galaxies (the satellites) which reside in groups
and clusters. We argued that, from the point of view of Halo Model 
analyses, it is interesting to ask whether the luminosities of the
BCGs retain any signature of their special nature. Our analysis shows
that, at the level of one-point distributions, the answer appears to 
be `no': 
BCG luminosities (insofar as they are accurately represented in the 
Mr20 catalog that we analysed) are consistent with being the statistical 
extremes of a universal luminosity function, which is well-described by 
\eqn{pgal-bernardi}.

We also analysed the luminosity functions of the satellites in
the Mr20 catalog. The behaviour of these distributions for different
values of group richness (i.e. number of members) $N$ is especially
interesting, since it potentially allows us to distinguish between two seemingly
reasonable but incompatible predictions of Halo Model analyses. On the
one hand, since the BCG luminosities are simply statistical extremes
of a universal galaxy luminosity distribution, it is reasonable to
expect that (a) the satellite luminosity function is just that given
by subtracting the BCGs from the universal one
(equation~\ref{app-psat}), and (b) the \emph{brightest} satellite
(BSG) luminosities are simply the \emph{second} brightest of $N$ draws
of the universal galaxy distribution. On the other hand, the HOD model 
predicts that (c) the satellite function is itself approximately
universal, and (d) the BSGs are statistical extremes of this
function. We showed that these two sets of predictions are
inequivalent in general.
In the present case, however, the large values of \Ngal\ that we consider prevent us from distinguishing between these two predictions, each of which provides an acceptable description of the data.

Additionally, we performed other tests to check the consistency of our
results. We showed that the statistics of the luminosity gap (the
difference in absolute magnitudes of the BSG and BCG) in the Mr20
catalog are consistent with the predictions of order statistics
(equation~\ref{pgapfix}). In Appendix~B we argued that statistical 
tests for the mean BCG luminosity based on fixed $N$ are to be 
preferred over those based on, say, fixed total group luminosity 
$L_{\rm tot}$, showing how the latter may lead to biased conclusions. 
We also emphasized that, in order to perform unbiased tests, 
it is crucial to use accurate descriptions of the bright tail of the 
galaxy luminosity function.

However, we showed that, at the level of two-point statistics, the 
BCG and BSG luminosity distributions are inconsistent with their 
having being drawn from a universal luminosity function (i.e., one 
that is the same for all groups).  We argued that because the 
luminosity mark correlation function differs significantly from unity 
on large scales (Figure~\ref{shuffledMarks}) the luminosity function 
of groups must depend on group properties (e.g. mass).  
This was not at all obvious from the (more traditional) one-point 
analysis of the group and BCG luminosity functions themselves.  

In work in progress, we show that the extreme values assumption together with a universal luminosity distribution \pgal\ 
predicts a halo mass dependent BCG luminosity function $g_1(L|m) = \sum_N p(N|m) g_1(L|N)$ if the distribution of \Ngal\ depends on mass $m$. This would have immediate consequences for implementations of the Halo Model such as the Halo Occupation Distribution (HOD, Zehavi \etal\ 2005), since such analyses would considerably simplify upon imposing the extreme value restriction on the BCG luminosities. However, we have already seen that the assumption of universality is inconsistent with the observed mark correlation signal on large scales. One might then wonder if the correct solution is to replace $p_{\rm gal}(L)\to p_{\rm gal}(L|m)$. In this case, the BCG luminosity function averaged over \Ngal\ is mass dependent because both $p_{\rm gal}$ (and hence $g_1(L|N,m)$) and $p(N|m)$ depend on $m$. In this case it is no longer obvious that the extreme values hypothesis should be tested at fixed \Ngal\ as we argued earlier. Formulating the problem cleanly in this case is work in progress.

Additionally, as discussed in the Introduction, this behaviour of the
BCG luminosities, in particular, the fact that their luminosity
functions are narrower than those of all galaxies, potentially allows
BCGs to be used for consistency checks of standard candle constraints
on the luminosity-distance relation.
As a final remark, we note that, while BCGs have brighter and narrower
luminosity distributions than randomly picked galaxies, the
distributions of the BSG luminosities are even narrower than those of
the BCGs. From the point of view of using them as standard candles, an
interesting trade-off then arises: while it is the BCG which is most 
easily observed out to large distances, the BSG is a more standard 
candle. It will be interesting to see to what extent these ideas can be
implemented in upcoming cluster surveys.

\section*{Acknowledgments}
We thank an anonymous referee for useful comments on an earlier version of the paper.

\appendix
\section{Prediction for the satellite luminosity distribution from
         extreme value statistics}\label{psatExstat}
If the luminosity distribution of galaxies \pgal\ is universal, and 
if the BCGs in groups containing $N$ members are described as the
brightest objects of $N$ independent draws from this distribution,
then the distribution of the satellites in such groups must have a 
specific dependence on $N$, as we show below.

Consider a sample of $\Cal{N}_{{\rm g}|N}^{>L_{\rm min}}$ galaxies
brighter than $L_{\rm min}$, residing in $\Cal{N}_{{\rm
    h}|N}^{>L_{\rm min}}$ groups (halos) which contain exactly
$N$ members each, so that $\Cal{N}_{{\rm g}|N}^{>L_{\rm min}} =
N \Cal{N}_{{\rm h}|N}^{>L_{\rm min}}$. Since \pgal\ is
universal, the number of these galaxies brighter than $L$, is
$\Cal{N}_{{\rm g}|N}^{>L} = \Cal{N}_{{\rm g}|N}^{>L_{\rm
    min}}\Pgal(>L)$. Similarly, the number of BCGs brighter than $L$
is $\Cal{N}_{{\rm BCG}|N}^{>L}= \Cal{N}_{{\rm h}|N}^{>L_{\rm
    min}} g_1(>L|N)$, where $g_1(>L|N) = 1 - \Pgal(<L)^{N}$ follows
from \eqn{g1cum}. 

The satellites are obtained by subtracting the BCGs from the full
sample, and the number of satellites brighter than $L$ is therefore
$\Cal{N}_{{\rm sat}|N}^{>L} = \Cal{N}_{{\rm g}|N}^{>L} -
\Cal{N}_{{\rm BCG}|N}^{>L} $, which tells us that
$\Psat(>L|N)$ is
\begin{align}
\Psat(>L|N) &= \frac{\Cal{N}_{{\rm sat}|N}^{>L}}{\Cal{N}_{{\rm
      sat}|N}^{>L_{\rm min}}} = \frac{\Cal{N}_{{\rm g}|N}^{>L} -
\Cal{N}_{{\rm BCG}|N}^{>L}}{\Cal{N}_{{\rm g}|N}^{>L_{\rm min}} -
\Cal{N}_{{\rm h}|N}^{>L_{\rm min}}}\nonumber\\
&=\frac{1}{N-1} \left( N\Pgal(>L) - g_1(>L|N)\right)
\nonumber\\ 
&= 1 - \frac{N\Pgal(<L) - \Pgal(<L)^{N}}{N-1}\nonumber\\ 
&= \Pgal(>L) - \frac{\Pgal(<L) - \Pgal(<L)^{N}}{N-1}\,,
\label{app-Psat}
\end{align}
which gives a satellite luminosity distribution
\be
p_{\rm sat}(L|N) =
\frac{1}{(1-1/N)}\pgal\left(1-\Pgal(<L)^{N-1}\right) \,.
\label{app-psat}
\ee
Notice that for large enough richness values $N$, 
 $1-1/N\to1$ and $\Pgal(<L)^{N-1}\to 0$, 
for all interesting values of $L$, and hence
$p_{\rm sat}(L|N\gg1)\to\pgal$.  
\fig{fig-psat} shows that averaging this distribution over the 
observed distribution of $N$ provides a good description of 
the distribution of satellite luminosities in the B06 group 
catalog, especially at large $N$. 

The extreme value statistics associated with this distribution yields 
\be
 g_{\rm 1sat}(<L|N) 
  = \left(\frac{N\Pgal(<L) - \Pgal(<L)^{N}}{N-1}\right)^{N-1},
 \label{g1sat}
\ee
which is clearly different from $g_2(<L|N)$ in the main text.
Although they are different for general $N$, it is easy to show that
\be
 g_{\rm 1sat}(<L|N) \to g_2(<L|N) \qquad {\rm as}\quad N\gg 1;
\ee
i.e., the distribution of the largest of $N-1$ picks from the 
satellite distribution does indeed tend to that of the second largest 
of $N$ picks from the full distribution.  We check this explicitly in 
Figure~\ref{fig-g1satavg-Mr20} of the main text.

\section{A shuffling-based test of extreme value statistics}\label{shuffled}

\begin{figure*}
 \centering
 \includegraphics[width=0.9\hsize]{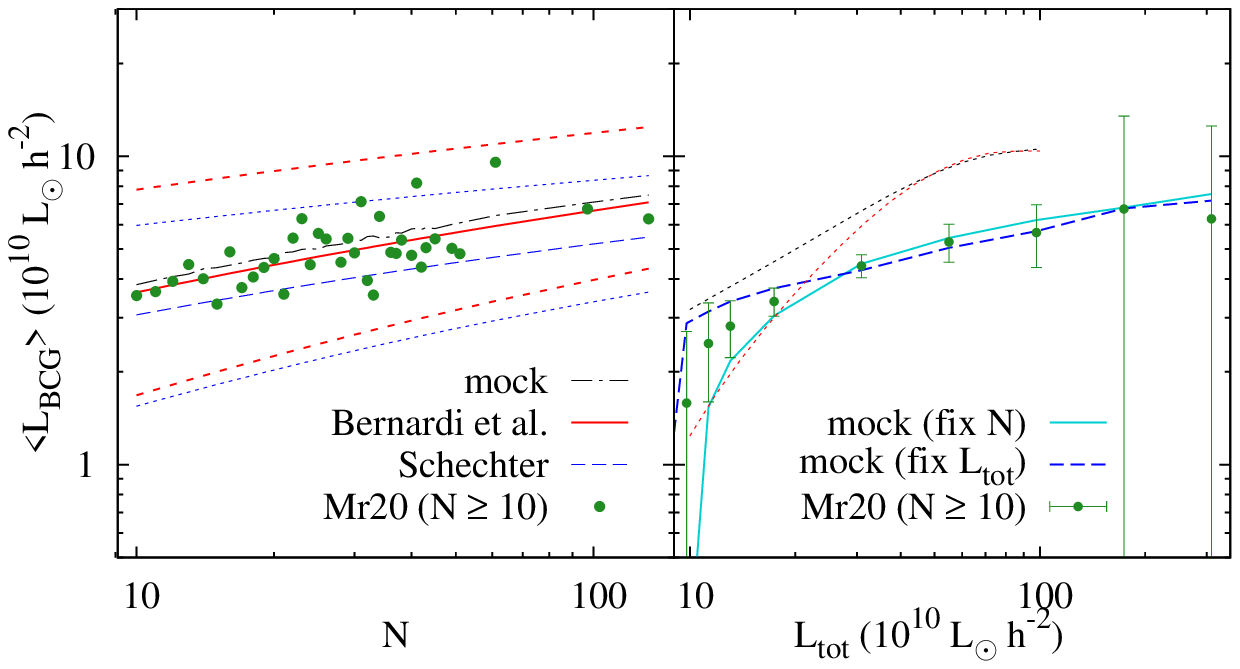}
 \caption{Mean value of $L_{\rm BCG}$ as a function of group size 
          \Ngal\ (left) and $L_{\rm tot}$ (right) in the B06 Mr20
          catalog (symbols). 
          \emph{(Left panel):} Dotted-dashed line shows the
          mean relation between $L_{\rm BCG}$ and \Ngal\ in 500 mock 
          catalogs obtained by scrambling the observed list of galaxy
          luminosities and constructing groups such that the observed
          distribution of \Ngal\ values is reproduced (see
          text). Solid red curve shows the mean prediction of extreme
          value theory based on \eqn{pgal-bernardi} with
          \eqn{params-bernardi}, while red 
          short-dashed curves indicate the corresponding region
          enclosing 95\% of probability for each \Ngal. The data,
          mocks and theory predictions are in excellent
          agreement. 
          Blue long-dashed and corresponding dotted curves show the
          mean and 95\% region of extreme value theory based on the
          Schechter form of Blanton \etal\ (2003)
          \eqn{params-blanton}, and emphasize that 
          using an inaccurate description of the bright tail of the
          galaxy luminosity function can lead to erroneous conclusions
          regarding, e.g., the statistical significance of outliers. 
          \emph{(Right panel:)} Solid line shows the mean relation
          between $L_{\rm BCG}$ and $L_{\rm tot}$ in the same mocks
          which gave the dotted-dashed curve of the left panel. The
          data agree very well with these mocks, although they seem to
          lie systematically slightly higher than the curve at small
          $L_{\rm tot}$. Dashed curve shows
          the same mean relation, but in mocks generated by fixing an
          observed $L_{\rm tot}$ value and randomly sampling the
          observed galaxy luminosity list until the total luminosity
          exceeds this $L_{\rm tot}$ (see text for more details). This
          procedure, in general, does not reproduce the observed
          distribution of \Ngal, and leads to biased conclusions
          regarding the mean relation.
          The dotted curves explicitly demonstrate this bias for 
          the exponential luminosity distribution
           $P(>L)={\rm e}^{-(L-L_{\rm min})/L_\ast}$ by
          plotting $\avg{L_{\rm BCG}|L_{\rm tot}}$ for two choices 
          of the distribution of richness $p(N)$ (see text).}
 \label{fig-shuffled}
\end{figure*}

In the main text, we argued that \eqn{pgal-bernardi}, with 
parameters from Bernardi \etal\ (2010) provided a good fit to the 
distribution of luminosities in the B06 group catalog.  Here, we 
provide additional tests of this assumption.  We also make two points: 
First, when one is interested in deriving constraints from the extreme 
tails of a distribution, then it is important to have an accurate 
description of the tails.  
Second, when testing extreme value or, more generally, order statistics, 
the test is best done keeping \Ngal\ fixed.  

A simple test of whether or not BCG luminosities are unusual compared 
to those of the other galaxies in the group is to make a mock catalog 
by scrambling the list of galaxy luminosities among the different groups.  
The BCG luminosity distribution in the mock catalog represents the 
expected distribution if BCGs were not otherwise special, and comparison 
with the BCGs in the original dataset indicates if they are indeed 
special.  In practice, we generate 500 realizations of the scrambled 
mock catalog, so as to produce a less noisy estimate of the expected 
extreme value distribution. 

The B06 groups are characterized by two numbers:  the total number 
of galaxies, \Ngal, and the total luminosity of the group $L_{\rm tot}$ 
(simply the sum of the luminosities of \Ngal\ group members).  When 
generating the mock catalog, we must decide whether to keep \Ngal\ or 
$L_{\rm tot}$ fixed.  Order statistics are clearly a function of \Ngal, 
so \Ngal\ is the natural variable.  However, one might be concerned that 
the luminosity of the BCG depends on `hidden' variables, such as halo 
mass.  If so, if one believes $L_{\rm tot}$ is a better indicator of 
this hidden variable, then one might wish to make mocks by holding it, 
rather than \Ngal, fixed.  We will show results of doing both, but note 
that the results in the main text strongly suggest that the test at 
fixed \Ngal\ is the prefered one.  

To generate a mock catalog, we first randomly scramble the observed
list of galaxy luminosities (which we take from groups with
$\Ngal\geq10$). We then run over the observed values of
\Ngal, sequentially picking \Ngal\ luminosities from the scrambled
list for each \Ngal, and store both the sum and the largest of these 
\Ngal\ picks. We then move on to the next value of \Ngal. This
procedure, by construction, reproduces the observed distribution of
\Ngal\ values. 

The dotted-dashed curve in the left hand panel of
Figure~\ref{fig-shuffled} shows the resulting correlation between
\Ngal\ and $L_{\rm BCG}$ in the mocks. This curve represents the
extreme value prediction for the $L_{\rm BCG}$, if BCGs are just the
extremes of the group-galaxy luminosity function.  The symbols show
the actual correlation in the Mr20 catalog (wherever possible, we
average over the BCG luminosities at fixed \Ngal).  We have
deliberately chosen to show the log of $L_{\rm BCG}$ because the
extreme value distribution has a long tail to large $L$ -- showing the
log brings the visual impression of the scatter around the mean value
closer to the correct one.

The solid curve shows the associated prediction for this correlation 
if we insert \eqn{pgal-bernardi} in \eqn{g1gal}.  It lies very close
to the dotted-dashed curve.  This is consistent with 
one of the points made in the main text: that the all-galaxy 
luminosity function, \eqn{pgal-bernardi} with \eqn{params-bernardi},
provides an excellent description of the Mr20 group-galaxy
luminosities. (The fact that it is systematically beneath the
dot-dashed curve is consistent with the fact that the galaxies in this
sample are slightly brighter than predicted by \pgal.)
The short dashed curves on either side of it indicate the region which
contains 95\% of the probability (at each \Ngal), also calculated from
\eqn{g1gal}. 
We note that none of the observed groups lie outside this
region.

The other dashed curve, and associated 95\% regions, shows the 
extreme-values prediction if we use the Schechter function of 
Blanton \etal\ (2003), \eqn{params-blanton}, instead.  This curve
falls below that of our mock catalogs.  Since a number of the Mr20
points lie outside its 95\% band, had one used this curve, rather than
performing the full shuffling procedure, one might have concluded that
the BCGs in the Mr20 catalog are inconsistent with the extreme values
hypothesis.   
However, this conclusion is unwarranted, because the Schechter
function does {\em not} describe the bright-end of the group-galaxy
distribution very well (see Figure~\ref{fig-pgal}).  Thus, the
exercise above illustrates the importance of using accurate measures 
of the tail, if one intends to use the tail to draw important
conclusions.   

The panel on the right shows the correlation between $L_{\rm BCG}$ 
and $L_{\rm tot}$ in the Mr20 catalog (symbols) and in the mock
catalogs which were used to produce the panel on the left (solid
curve). The agreement is very good, at least at large $L_{\rm tot}$,
again indicating that BCG luminosities are consistent with just being
statistical extremes.   
We have also checked that the mean $L_{\rm tot}$ at fixed \Ngal\ is
close to being proportional to \Ngal\ for this catalog, with the best
fit power law relation being $L_{\rm tot} = 1.66\times 10^{10} \Lsun
h^{-2} \Ngal^{0.96}$. 

The dashed curve, which lies above the measurements at small 
$L_{\rm tot}$ but close to the solid curve at larger $L_{\rm tot}$ was
obtained by generating mock catalogs following a slightly different
procedure -- one that is essentially the same as that recently used by
Lin \etal\ (2010) in their analysis of BCG luminosities in another
group  catalog.   
In this case, for each observed $L_{\rm tot}$, we continue to draw 
luminosities from the group galaxy distribution until the sum of 
these luminosities first exceeds $L_{\rm tot}$.  We then compare this 
value and the one preceding it to the observed value, and assign the 
closer of the two to the mock group.  We also determine the luminosity  
of the mock BCG from the list of luminosities which contribute to the 
group's luminosity, before moving on to the next value of $L_{\rm
  tot}$. To obtain smoother results, we run over the observed
distribution of  $L_{\rm tot}$ values 500 times.   

In this case, there is no guarantee that the mock catalog which 
results has the correct distribution of \Ngal. 
Since extreme value statistics are explicitly a function of the number 
of draws \Ngal\ (e.g. equations~\ref{g1gal} and~\ref{g2gal}), it seems  
unlikely that this procedure can yield a fair test of these statistics.  
Indeed, the difference between the dashed and solid curves in this panel 
is a measure of the bias in this particular test, since the solid line 
is associated with mock catalogs that, as shown in the panel on the 
left, are in excellent agreement with the expected extreme value 
distribution.  

This is easier to appreciate with a simpler example.  
When the universal function $p(L)$ is an exponential,
 $P(>L) = {\rm e}^{-(L-L_{\rm  min})/L_\ast}$, 
then the mean BCG luminosity at fixed $L_{\rm tot}$ and $N$ takes 
the form 
\be
\avg{L_{\rm BCG}|L_{\rm tot},N} = L_{\rm min} + \frac{H_N}{N}(L_{\rm tot}-NL_{\rm min}) 
\label{lbcggivenltot}
\ee
(defined for $L_{\rm tot}\geq NL_{\rm min}$), where
$H_N=\sum_{r=1}^N(1/r)$ is a harmonic number. The mean BCG luminosity
is then
 $\avg{L_{\rm BCG}|L_{\rm tot}} = \sum_N \avg{L_{\rm BCG}|L_{\rm tot},N}
   p(N|L_{\rm tot})$. 
The distribution $p(N|L_{\rm tot})$ is constructed as
 $p(N|L_{\rm tot})=p(L_{\rm tot}|N)p(N)/\sum_{N'} p(L_{\rm tot}|N')p(N')$, 
where 
\be
L_\ast p(L_{\rm tot}|N) = \frac{{\rm e}^{(L_{\rm tot}-NL_{\rm
    min})/L_\ast}}{(N-1)!}\left(\frac{L_{\rm tot}-NL_{\rm
    min}}{L_\ast}\right)^{N-1} \,,
\notag
\ee
which is also defined for $L_{\rm tot}-NL_{\rm min}>0$ and is zero
otherwise. 

One therefore needs to choose a ``prior'' distribution $p(N)$, and 
the dotted curves in right panel show $\avg{L_{\rm BCG}|L_{\rm tot}}$ for 
the exponential luminosity distribution when $p(N)$ is the one observed 
in the data (lower, red) and in the mocks described above which kept 
$L_{\rm tot}$ fixed (upper, black). We see that the difference between 
the curves has the same sense as that for the mocks based on the 
actual luminosity distribution. This explicitly demonstrates the 
bias introduced by using the wrong distribution of $N$.

We conclude that the BCG luminosities in the Mr20 group catalog are 
consistent with being the statistical extremes of the group-galaxy 
luminosity function -- the latter being very well described by the 
all-galaxy luminosity function.

\section{The HOD model of Zehavi \etal\ (2011)}
\label{zehaviHOD}
The HOD implementation of Zehavi \etal\ (2011) parametrizes the mean
number of galaxies $\avg{N|>L,m}$ brighter than some threshold $L$ in
a halo of mass $m$ using
\be
\avg{N|>L,m} = f_{\rm cen}(>L,m) \left(1+\bar N_{\rm
  sat}(>L,m)\right)\,, 
\label{hod-meanN}
\ee
where
\begin{align}
f_{\rm cen}(>L,m) &= \frac12\left(1 + {\rm erf}\left(\frac{\log m-\log
  M_{\rm min}(L)}{\sig_{\log M}(L)}\right) \right)\,,
\label{hod-fcen}\\
\bar N_{\rm sat}(>L,m) &= \left(
\frac{m-M_0(L)}{M_1^\prime(L)}\right)^{\alpha(L)} \,.
\label{hod-Nsat}
\end{align}
The function $f_{\rm cen}(>L,m)$ is interpreted as the number of 
centrals, and the product of $f_{\rm cen}(>L,m)$ and $\bar N_{\rm
  sat}(>L,m)$ as the mean number of satellites in the $m$-halo. It is
natural to interpret these functions in terms of conditional
probabilities, with $f_{\rm cen}(>L,m)$ being the fraction of
$m$-halos that host a central galaxy (note that it varies between 0
and 1 by definition) and $\bar N_{\rm sat}(>L,m)$ being the mean
number of satellites in halos that host a central. The distribution of
the number of satellites brighter than $L$  in an $m$-halo which hosts a central is 
assumed to be Poisson with mean $\bar N_{\rm sat}(>L,m)$. 

The values of the parameters $M_{\rm min},\sig_{\log
  M},M_0,M_1^\prime$ and $\alpha$ at various thresholds are then fit
by comparing a Halo Model calculation of the projected $2$-point
correlation function to measurements of luminosity dependent
clustering in SDSS data. The results are in Table 3 of Zehavi
\etal\ (2011), of which we use information for the 5 threshold values
brighter than $-19.9$. The HOD prediction for the satellite luminosity
distribution is 
\be
P_{\rm sat}(>L|m,L_{\rm min}) = \frac{\bar N_{\rm sat}(>L,m)}{\bar N_{\rm
    sat}(>L_{\rm min},m)}\,.
\label{hod-Psatm}
\ee
Since this depends on mass, obtaining predictions for the distribution at
fixed \Nsat\ requires an average over the mass. Following Skibba \etal\
(2007) we define the number density of groups with \Nsat\ satellites 
\be
n_{\rm grp}(\Nsat|L_{\rm min}) = \int dm \frac{dn}{dm} f_{\rm
  cen}(>L_{\rm min},m) p(\Nsat|m,L_{\rm min})\,,
\label{hod-ngrp}
\ee
where $dn/dm$ is the halo mass function and $p(\Nsat|m,L_{\rm min})$ 
is a Poisson distribution with mean $\bar N_{\rm sat}(>L_{\rm
  min},m)$. For brevity we drop the reference to $L_{\rm min}$ in what
follows. 

The satellite luminosity distribution at fixed \Nsat\ is then given by
\begin{align}
P_{\rm sat}(>L|\Nsat) &= \int dm \frac{dn}{dm} \frac{f_{\rm
    cen}(>L_{\rm min},m)}{n_{\rm grp}(\Nsat)} \nonumber\\
&\ph{dmdn/dm}
\times p(\Nsat|m) P_{\rm sat}(>L|m)\,.
\label{hod-PsatN}
\end{align}
This function is then used to compute an extreme value distribution
$g_{1\rm sat}(L|\Nsat)$, which is identical in form to \eqn{g1gal},
with $\Ngal\to\Nsat$ and $\pgal\to p_{\rm sat}(L|\Nsat)=-\partial
P_{\rm sat}(>L|\Nsat)/\partial L$. To be consistent, the
prediction for $\Nsat\geq N_{\rm min}$ must be averaged using the
distribution $p(\Nsat) = n_{\rm grp}(\Nsat)/\bar n_{\rm grp}$ where
$\bar n_{\rm grp} = \int dm dn/dm f_{\rm cen}(>L_{\rm min},m)$. In
other words, 
\be
g_{1\rm sat}(L)=\frac{\sum_{\Nsat\geq N_{\rm min}}g_{1\rm
    sat}(L|\Nsat) n_{\rm grp}(\Nsat)}{\sum_{\Nsat\geq N_{\rm min}}
  n_{\rm grp}(\Nsat)}\,.
\label{hod-g1avg}
\ee
Note that computing $g_{1\rm sat}(L|\Nsat)$ requires taking a
derivative of $P_{\rm sat}(>L|\Nsat)$. To ensure smooth results, in
practice we first fit simple monotonic forms to the values of the
various HOD parameters in Table 3 of Zehavi \etal\ For the mass
function we use the analytical approximation of Sheth \& Tormen
(1999). We have checked that the luminosity dependent clustering
predicted by our smooth fits to the Zehavi \etal\ values is close to
the measurements in their Figure 10 and Table 8. We then use $L_{\rm
  min} = 0.747\times 10^{10}\Lsun h^{-2}$ and $N_{\rm min}=10$, $15$ 
to make our predictions. These are shown as the dashed red curves in
\fig{fig-g1satavg-Mr20}.

\label{lastpage}

\end{document}